\documentclass[12pt]{iopart}
\usepackage{amsfonts}  
\usepackage{amssymb}
\usepackage{graphicx}   
\begin{document}

\title
{Applications of a constrained mechanics methodology in economics}

\author{Jitka Janov\'a  $^{1\, 2}$ }

\address{$^1$ Department of Theoretical Physics and
Astrophysics, Faculty of Science, Masaryk University, Kotl\' a\v
rsk\' a 2, 611 37 Brno, Czech Republic}

 \address{$^2$ Department of Statistics and Operation Analysis,
 Faculty of Business and Economics,
Mendel University in Brno, Zem\v ed\v elsk\'a 1,
613 00 Brno,
Czech Republic}

\ead{janova@mendelu.cz}   

\begin{abstract}
The paper presents instructive   interdisciplinary applications of  constrained mechanics calculus in economics on a level appropriate for the undergraduate  physics education. The aim of the paper  is:
\begin{enumerate}
 \item  to  meet the demand for illustrative examples suitable for presenting the background of the highly  expanding research field of econophysics even on the undergraduate level and
 \item to enable the students to understand deeper the  principles and methods  routinely used in mechanics by looking at the well known methodology from the different perspective of economics.
\end{enumerate}

Two   constrained dynamic economic problems  are presented using the economic terminology in an intuitive way. 
First, the Phillips model of business cycle is presented  as a system of forced oscillations  and the general  problem of two interacting economies is solved by the nonholonomic dynamics approach. 
Second, the Cass-Koopmans-Ramsey model of economical growth  is solved as a variational problem with a velocity dependent constraint using the vakonomic approach. 
The specifics of the solution interpretation in economics compared to mechanics is discussed in detail, a discussion of the nonholonomic and vakonomic approaches to constrained problems in mechanics and economics is provided  and an economic  interpretation  of the Lagrange multipliers (possibly surprising for the students of physics) is carefully explained.

The paper can be used by the undergraduate students of physics  interested in  interdisciplinary physics applications  to get in touch with current scientific approach to economics based on a physical background or by university teachers as an attractive  supplement to the classical mechanics lessons. 


\end{abstract}

\maketitle

\section{Introduction}
The   economic science has been influenced by physical concepts from its very beginning.\footnote{ Let us mention  Daniel Bernoulli, as an example, who was the originator of utility-based preferences, or one of the founders of neoclassical economic theory, Irving Fisher, who was originally trained in physics under   J. W. Gibbs. }.
Although it is for contributions to physics and mathematics that  Newton is celebrated,  the Newtonian principles, formulated at the end of the seventeenth century, have  powerfully  influenced most branches of science -  economics undoubtedly.  
 Newton's mechanics brought the doctrine of scientific determinism, the principle that all events are the inescapable results of preceding causes, for  which (until the work of Planck and Einstein in twentieth century) the scientists tend  to think of nature as a mechanical device whose behavior could be revealed by observation, experimentation, measurement and calculation. The idea of nature  governed by natural laws dominated the new world order and many scholars have presumed human behavior and economics to be governed by such  laws as well (see \cite{Canterbery}). 

  As the economic thinking  were  developing through the centuries, the possibility that economic science can be inspired by physics was  continuously debated. Nevertheless,  the newly built  physical theories and methodologies were repeatedly applied to economic problems by scholars trying to capture observed  economic behavior.  As a well developed branch of physics perfectly equipped with mathematical apparatus, mechanics has permanently served an inspiration for theoretical  constructions in economics (for detailed discussion see \cite{GG}).
 
Looking for  recently arisen  intersections of physics and economics one  arrives to {\it econophysics} which  describes the phenomena of development and dynamics of economic systems by using  a strictly  physically motivated methodology. The official birth of the term "econophysics" dates back to  a paper by H. Stanley  in 1996 \cite{Stanley}. Currently, mainly the  applications of  statistical physics and nonlinear dynamics are considered a core of econophysics (see \cite{Schulz}), but 
in broader context,  econophysics can be considered  an  interdisciplinary research field applying  theories and methods originally developed by physicists in order to solve problems in economics (for approach employing solely the  classical mechanics methods  see  \cite{Kitov}).   

 Econophysics is a modern, quickly expanding  interdisciplinary branch of science, which has  already been transferred from the area of purely scholar interest into the real world  including  the establishment of graduate and postgraduate university study programs. Although econophysics has become a part of university physics education,  the teaching aids are still being developed. Since the core of  econophysics lessons requires  passing advanced physics courses, very little is available for undergraduate students to satisfy their curiosity about this "econophysics fashion". In this paper we present instructive examples of how the methodology of constrained dynamic systems, commonly used in classical mechanics, can  be used for solving  economic problems and, in this way, we  supply  instruments to present and demonstrate this interdisciplinary topic on the  undergraduate level of physics education. For recent educational contributions dealing with econophysics  (but lacking concrete examples or inappropriate for undergraduates) see \cite{Schinkus}, \cite {Walstad}, \cite{Stauffer}, \cite{Mostardinha}.
 Being intended for the students of physics without preceding  knowledge of economic theories, the examples given in the paper are based on intuitive economic terminology and  supplemented by easy-to-read explanations and interpretations of the economical background. 

Apart from the illustrative potential of the examples, the look at standard methodology procedures used in mechanics (such as formulation of the equations of motion or handling the constrained systems) from a completely different point of view enables the reader to reach a better understanding of the physical background  of these principles - a point which is  often replaced  by  the calculation routine otherwise. 

The paper can be used by  undergraduate students of physics interested in solving  interdisciplinary applications to get an idea about employing physical apparatus in economic problems. 
 Moreover, the examples provided in this paper can be utilized by university teachers as an attractive  supplement of traditional mechanics courses on the undergraduate level or as a motivation inviting the students to enroll for advanced interdisciplinary courses. 
 



In Sec. 2 the classical Phillips model of business cycle is solved via the physical model of forced oscillations under friction: in Sec.  2.1 the economic model is briefly developed   in an intuitive way,  in Sec 2.2 the problem of two interacting economies is stated and an appropriate, physically motivated model of a nonholonomic system is used for its  solution and in Sec. 2.3 we  demonstrate  that the results obtained are in good qualitative correspondence with the observed behavior of two economies.
 In Sec. 3,  the Cass-Koopmans-Ramsey model of economic growth is investigated by the calculus of variations. It is shown, that the Lagrangian has the meaning of the overall (current value) utility in economy and the existing connection  among the economic quantities  is modeled by a velocity-dependent constraint. The vakonomic approach is used for solving the system and the examples of a typical economic solution procedure and the interpretation of the solution are presented and discussed with respect to the standard procedures used in mechanics.  A discussion of the difference between  the nonholonomic and vakonomic approach together with the possibilities of their application in economics and mechanics is provided in Appendix 1. A special paragraph is devoted to
an interesting economic interpretation of the  Lagrange multipliers and   an additional example and comments on this topic are given in Appendix 2.

\section{Physically motivated business cycle description} 

Business cycle (or economic cycle) refers to economy-wide fluctuations in economic activity over several months or years. These fluctuations occur around a long-term growth trend, and typically involve  periods of  rapid economic growth (an expansion), and periods of relative stagnation or decline (a contraction or recession, see e.g. \cite{Sullivan}). Business cycles exist in economy of any country or region (e.g. we can detect business cycles in Spanish economy or business cycles of European Union economy). Understanding how this cycles come into existence and what are their determinants, is crucial for making good economical policies in a particular country/region since the level of economic activity is linked with the  standard of living in a country. Therefore mathematical models of business cycles form an important and intensively studied topic in the science of economics\footnote{Note that currently there is a number of different mathematical models of business cycle and still new are being  developed. Generally in economic science, for each economic phenomenon (such as   business cycle), there exist  several competing approaches   that use different assumptions and   mathematical apparatus for building the model. Whatever  the details of these approaches, all of them   aim to  follow the current and forecast the future  behavior of the economic quantities as precisely as possible.  Interesting from the physical point of view is  that there is no unique "correct" model  in almost any economic discipline. There are many models for every phenomenon, each having its  advantages and disadvantages and each  corresponding with reality to an own extent.}.

The business cycle is  described via the time development of gross domestic product ($GDP$) which refers to the market value of all final goods and services produced within a country in a given period. The term "gross" refers to particular methodology of enumerating $GDP$ and in economic models we speak simply about product (denoting it $Y$).  Hence, the product $Y=Y(t)$ provides us with the information on how much goods and services is produced (and it is assumed that this amount is also sold) in the economy (of a country/region) in each time period. When trying to  express   $Y=Y(t)$ quantitatively, the economists face the problem how to find the representative function. Unlike in mechanics where the time development of the system is  uniquely determined by forces and torques acting on the system,  no such unequivocal approach exists in economics. 
Economists state that the characteristics playing a principal role in forming the business cycle are:
\begin{itemize} 
\item investments  $I= I(t)$ - the purchased amount  of goods which are not consumed but are to be used for future production  (e.g. purchases of new machines and buildings intended to be used in a production process),
\item consumption of households $C=C(t)$ - the amount of money spent by households for goods and services, 
\item total demand for goods and services $Z=Z(t)$ - the amount of goods and services intended to be purchased in the economy. Let us explain more carefully the concept of total demand using the illustration of money/goods and services flows in an economy (see Fig. \ref{picture-economy}):
\begin{figure}[th]
\begin{center}
\scalebox{0.5}{\includegraphics{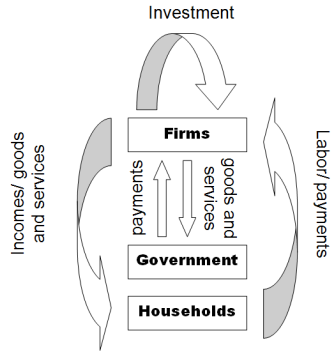}}
\caption{\label{picture-economy} The illustration of money/goods and services flows in an economy}
\end{center}
\end{figure}
There are three "players" in an "economy game": households that  consume goods and services  spending  the  money earned by working in firms, firms who are renting  labor from workers to produce goods and services and a government that  purchases both  goods and services. Hence, the total purchases of goods and services  (i.e. demand $Z$) is composed of the household consumption, firm investments and government purchases of goods and services.
\item autonomous expenditures $A=A(t)$ - expenditures independent from the total income in economy (total income is the amount of money earned by the individuals in economy), i.e. whatever the economic performance of the economy is, these money will be spent (e.g. government expenditures to keep the roads functional).  
 Note that product $Y$  (which is expressed in monetary units) is identified with total income in the economy\footnote{   The product $Y$ is the total amount of goods and services produced in economy. But what was produced is to be sold and all the money are disbursed to the players in economy game (e.g. the employees are paid a wage, the material or  semi-finished products must be bought from firms - who must paid their employees etc.-  
and the surpluses of corporations and entrepreneurs come as income to individuals in the economy). Hence, what was produced in economy transforms into the total income in economy.}.
 
\end{itemize}

\subsection{The Phillips model of business cycle}

 In  the Phillips model   of business cycle (developed in 1954), which from the  physicist's point of view is interesting by the methodology used,  the product function $Y(t)$ is deduced from the observed and presumed relationships between the above mentioned characteristics of the economy (see e.g. \cite{Allen}): 
\begin{enumerate}
\item The total demand in economy $Z(t)$ is composed of the consumption of households $C=(1-s)Y$, $s=const$\footnote{The individuals divide their income between consumption (spending their money for goods and services) and saving (leaving the money in banks). So for the total income in economy (which can be denoted also $Y$ since it has the same value as product) we can write $Y=C+S$, where savings $S=sY$ and consumption $C=(1-s)Y$ are given by the income (product) $Y$ and constant value $s, \, 0 \leq s \leq 1$, which is a given characteristic of an economy. 
},  investments purchases  $I$ and the autonomous expenditures $A$ (we will consider it to be a given constant of government spending):
\begin{equation}
Z=(1-s)Y+I+A,\label{ma2}
\end{equation}

\item The demand in economy can be satisfied only via the product of the economy (since we do not take into account other economies).  Hence, ideally the demand $Z$ should be equal to supply (product) $Y$ in any time. But it is assumed (and more realistic) that the product $Y$ is reacting to the demand  with a delay. For example if there is a demand for some fashionable consumer article and there is a lack of it in shops, the producers can react and increase its production, but this will take some time. Thus the supply of the article reflects the past demand. In this way also total product $Y$ "persuades" the total demand $Z$ in time   and, particularly in Phillips model it is assumed that the dynamics of product is given by
\begin{equation}
\frac{{\rm d}Y(t)}{{\rm d} t}=\lambda (Z-Y),\label{ma3}
\end{equation}
where $\lambda$ is a positive  constant.
\item It is assumed that the potential investment $\tilde I(t)$ is depending on the  product change in time $\tilde I(t)=v\frac{{\rm d}Y(t)}{{\rm d} t}$ ($v$ being  constant), 
 i.e.  the investments are needed only if we want to increase production.
For example if we want to produce more products in a factory, we will need more inventories, machines, etc. which generates the spendings falling within the "investments" in the economic language.
  But the true  investment $I(t)$ is delayed from the potential one, i.e. in reality the investments do not react on the change in product immediately (in our example we can imagine, that we begin product more products using more extensively the machines we already have and the new machines will be  bought later).  The change of true investments in time is then supposed to be  proportional to the difference $(\tilde I(t)-I(t))$, i.e.
\begin{equation}
\frac{{\rm d}I}{{\rm d} t}=\kappa\left(v\frac{{\rm d}Y(t)}{{\rm d} t}-I\right), \label{ma1}
\end{equation}
where $\kappa$ denotes a positive constant. 

\end{enumerate}
Now the above mentioned equations describe the model of how business cycle come into existence:  the product is delayed from demand, and investment purchases are delayed  from their  immediate need. These two factors generate in Phillips model the oscillations in time development of product $Y$\footnote{

 The sources of the periodic oscillations observed in time development of product $Y$ still are the matter of investigation.  A modern  theory of {\it real business cycles} suggests that the fluctuations of product  can be to a large extent accounted for by real  shocks (examples of such shocks include innovations, bad weather, quick oil price increase, stricter environmental regulations, etc.) which appear more or less periodically. The general gist is that something occurs that directly changes  the decisions of workers and firms about what they buy and produce and thus eventually affect the product $Y$. 

Note then that  Phillips model represents only one of many possible reasoning for the product oscillations.}. 
 Combining equations (\ref{ma2}-\ref{ma1}) and eliminating $Z$ and $I$, we obtain a second-order differential equation for the unknown variable $Y$ which represents the business cycle:
\begin{equation}
\ddot Y+a\dot Y+b Y=P,\label{nevazrce1}\label{rov1}
\end{equation}
where
$$a=\lambda s+\kappa(1-\lambda v)$$
$$b=\kappa\lambda s$$
$$P=\kappa \lambda A$$
are constant values.
Equation (\ref{rov1}) reminds us of forced oscillations under friction where $b=\omega_0^2$ is the square of the free oscillations frequency and $P$ is the external force acting on the system - generally it is a given function of time (this is the case also in Phillips model once the autonomous expenditure $A$ is not constant). 
 In physics, the term  $a\dot Y$, $a>0$, represents the damping   arising  when the surrounding medium exerts a resistance. In an economic system  this term may cause both damped and explosive oscillations depending on the sign of $a$.

\begin{figure}[th]
\begin{center}
\scalebox{0.5}{\includegraphics{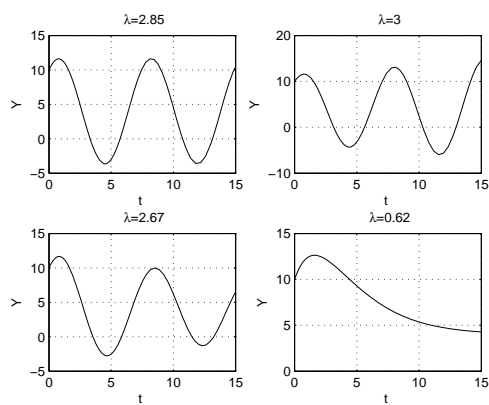}}
\caption{\label{unconstr} The solution of an unconstrained system; top-left: harmonic oscillations, top-right: explosive oscillations, bottom-left: damped oscillations, bottom-right:critical damping }
\end{center}
\end{figure}

Solving the dynamical equation (\ref{rov1}) for our economic problem,  we can obtain harmonic oscillations, critical damping, damped or even explosive oscillation according to given parameters of the economic system. Given all the other parameters, by changing  the constant $\lambda$ (which can be found in each of the constants $a,\,b,\,P$) we can obtain all  types mentioned of time development of the system. The solution of (\ref{nevazrce1}) for several values of $\lambda$ can be found in Fig. \ref{unconstr}\footnote{ The values of the parameters where chosen according to   \cite{Allen}:
 $  \kappa=1,\, A=1,\,  s=0.25,\,v=0.6$ and the initial conditions were set as $Y(0)=10 $
, $\dot Y(0)= 4$.}. 
Note that we do not mention  concrete units in the graphs, since the results are derived for  artificial parameters and for the purpose of demonstrating the behavior of  the product $Y$.  Visualizing "non-specified" graphs is a standard means of qualitative presentation of economic behavior. But if needed, one can scale the product $Y$  in thousands  of USD  and the time axis  in years. 

The manner in which  the product $Y$  is developing in time (harmonic, explosive, damped oscillations or  the critical damping) depends on the particular choice of parameters $\kappa,\, \lambda,\, s,\, v,\,A$. Given  the rest of parameters, the intervals of $\lambda$-values corresponding to each possible solution in Fig. \ref{unconstr} can be derived from a  general solution of the dynamical equation (\ref{rov1}). Analyzing these intervals, the economists conclude that oscillations,  will occur if the time lag of product reaction to the demand does not differ significantly from  the time lag of the reaction of investment to product change. The damped (explosive) oscillations will occur if the time lag of production to the demand will be significantly higher (lower) than the time lag of the reaction of  investments to product change.  
 These explanations  following from the mathematical model answer the question of how the business cycle come to existence only to some extent.   As we have mentioned there are many other approaches to business cycle modeling and always there  may be a debate on the   explanations and reasoning of each particular model. 

It is worth noting  that the final "equation of motion"  (\ref{rov1}) for the economic system  (we will call it  {\it a dynamical equation} when talking about economic systems)  was obtained based on  assumptions (\ref{ma2}-\ref{ma1}) about the dynamic relationships among the economic quantities.  But there is no mention about the Lagrangian of the economic system\footnote{Note that in next paragraph we will see that the variational concept  has also been adopted by quantitative economists in other models.} so as to  generate the dynamic equation (\ref{rov1}),  nor are the terms  in (\ref{rov1}) interpreted as some economic "forces". Thus, although the resulting dynamical equation is well known in physics,  the method of  obtaining it lacks the systematic spirit of physical methodology (remember Lagrangian approach or Newton equations in  mechanics.) 

Hence, through  certain assumptions Phillips model arrived to the description of oscillating behavior of product by the differential equation of second order. The ideal result of this model would be that once known the constants of the economy ($\lambda,\nu, s, \kappa, A$) one could predict the future oscillations of $Y$.  Practically, the political authorities would prefer only steady economical growth (i.e  growth of the product), which generates low unemployment, increasing wages and overall increasing life standard of  people. Oscillations around this simple growth trend are disturbing  and at least knowing how they come into existence or better, what they will look like next period, would bring important support to the policy makers.  But as we have mentioned before, such predictions and even the understanding of the oscillation phenomena are not completed and are the subject of ongoing research. 

Let us conclude the description of Phillips model of business cycle by the note that using the well-known and  developed physical model (the  harmonic oscillations) to describe  observed economic behavior (product time development)   was a natural  initial  approach in  economics. Although the current  philosophy of quantitative economics diverges fundamentally from simply adopting the existing physical models for the description of economic systems, the physical methodology itself is still in the focus of  economists. Accordingly, the  harmonic-oscillations-based methodology is still a common approach used for the description of economic  systems nowadays (for a recent application see e.g.\cite{Ataullah}). No matter how historical and simple the approach is, the Phillips model is able to produce at least qualitatively correct solutions (in the sense of  correspondence with the observed behavior) as we shall see in Sec. 2.2. 

\subsection{Two interacting economies }

In the preceding model we considered an isolated economy which had no contact with any other economy. In reality,  economies are interacting  through mutual purchases and money transfers. As typical in economics, a number of models were developed for this more realistic assumption. The most common is the model where, when describing a single economy, we introduce another "player" (apart from households, firms and government) representing external economies influencing the  economy modeled.  But let us continue to develop a  different, possibly more general, model employing the physical description  initialized in the preceding section.

Let us consider economies of two countries which are "interacting" through purchases and money transfers, i.e. part of the goods and services in an economy is purchased in the another. We can assume that  the product  in one economy will be affected besides other things also  by the  product of the other economy (for example if in Germany decreases the total product, i.e. there is less production of goods and services, then the demand of German firms for the sub-products and services imported from Czech Republic decreases  and this will naturally affect negatively the product of Czech Republic. And, reciprocally,  the less product in Czech Republic means lower income of households and lower consumption of all goods and services, among all those more expensive e.g.  those imported from Germany, so the lower demand in Czech Republic will affect -to some extent- the German product). 

 Assume that, if there was no such interaction between the economies, each of them  could be described in terms of equation (\ref{rov1}). Hence, we can describe two economies without interaction as a system with two degrees of freedom using the dynamical equations:

\begin{eqnarray}
\ddot Y_1+a_1\dot Y_1+b_1 Y_1&=&P_1,\label{nev1}\\
\ddot Y_2+a_2\dot Y_2+b_2 Y_2&=&P_2,\label{nev2}
\end{eqnarray}
with constants
$$a_i=\lambda_i s_i+\kappa_i(1-\lambda_i v_i),$$
$$b_i=\kappa_i\lambda_i s_i,$$
$$P_i=\kappa_i \lambda_i A_i,$$
where $i=1,2 $ labels the economies.
Following the physically motivated design of the Phillips model, we will formulate the interaction between the economies as a  constraint binding the two coordinates ($Y_1,\,Y_2$). In physics, the constraints  are used if the real forces ensuring the observed behavior are not known or are uneasy to  quantify. In a constrained system then, the constraint forces arise which ensure the prescribed behavior and have the meaning of real physical forces acting on the system (remember e.g. the  rolling of the cylinder without slipping in mechanics where the  constraint forces have the meaning of  forces and torques stemming from  interaction of the cylinder with underlay, see \cite{valce} or \cite{snakeboard}). Although we have already mentioned that no "forces" are defined in economics, the constraints seem to be an appropriate  methodology since  they make it possible  to describe  the really observed behavior whose "dynamical generator" is not known. 
The relationship between the two economies can be stated generally as a nonholonomic constraint. For our artificial example, let us assume this constraint to be  
\begin{equation}
\dot Y_2= k\dot Y_1+ \alpha Y_1+ \beta Y_2,
\label{vazba1}
\end{equation}
where $k, \alpha,\beta $ are constants. 
 Since the constraint is linear in velocities and the model itself is  physically motivated, we will treat the constrained system in a way  common in mechanics - we will use the nonholonomic approach (see e.g. \cite{Zampieri} or \cite{valce2}). 
Then, the dynamical equations for our constrained system take  the form:
\begin{eqnarray}
\ddot Y_1+a_1\dot Y_1+b_1 Y_1&=&P_1-\mu k,\label{deform1}\\
\ddot Y_2+a_2\dot Y_2+b_2 Y_2&=&P_2+\mu\label{deform2},
\end{eqnarray}
which, together with constraint (\ref{vazba1}), yields 3 equations for 3 unknown variables $(Y_1, Y_2, \mu)$, where $\mu$ is the Lagrange multiplier. Eliminating the Lagrange multiplier from   (\ref{deform1}-\ref{deform2})  and substituting from the constraint (\ref{vazba1}), we obtain reduced equations of the constrained system:
\begin{eqnarray}
(1+k^2)\ddot Y_1+A_1\dot Y_1 ++B_{11}Y_1+B_{12} Y_2-P_1-P_2&=&0,\label{redrov1}\\
\dot Y_2- k\dot Y_1+ \alpha Y_1+ \beta Y_2&=&0,\label{redrov2}
\end{eqnarray}
where
\begin{eqnarray}
A_1&=& a_1+a_2k^2+\alpha k+\beta k^2, \nonumber\\
B_{11}&=& b_1+\alpha a_2 k+ \alpha\beta k,\nonumber \\
B_{12}&=& \beta a_2 k+k b_2+k\beta^2. \nonumber 
\end{eqnarray}

\subsection{Results and discussion}
 Solving the constrained system (\ref{nev1}-\ref{vazba1}) for three different sets of parameters $k,\,\alpha ,\,\beta$ in (\ref{vazba1}):
\begin{tabbing}
Constraint A $\alpha=-0.1$\=  $\alpha=-0.1$ \= $\beta=-0.2$\kill
Constraint A:	\>	$\alpha=0$ \> $\beta=0$ \\
Constraint B:		\> $\alpha=0.4$ \> $\beta=0.1$ \\
Constraint C:	\>	$\alpha=-0.1$ \> $\beta=-0.2$ 
\end{tabbing}
  \begin{figure}[th]
\begin{center}
\scalebox{0.5}{\includegraphics{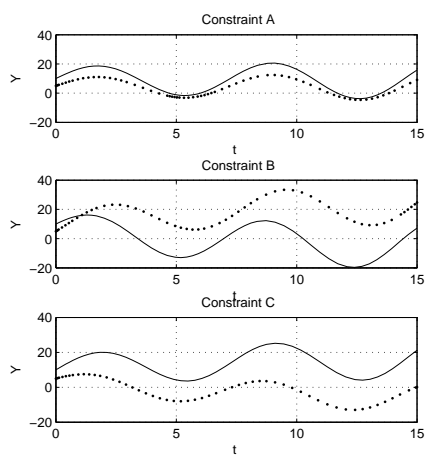}}
\caption{\label{constr} The solution of a  constrained system for a particular setting of  constraint parameters}
\end{center}
\end{figure} 
 we obtain the solutions  in Fig. \ref{constr}\footnote{The parameters used for our example were chosen  according to \cite{Allen}
$k=0.7,\, \lambda_1=3,\, \lambda_2=3,\, \kappa_1=1,\,\kappa_2=1,\, A_1=2,\,  A_2=1,\, s_1=0.25,\, s_2=0.25,\, v_1=0.6,\, v_2=0.6$. Hence, the economies are quite similar except  the autonomous expenditure $A$.
    The initial conditions for our problem were the following: 
$Y_1(0)=10$, $Y_2(0)=5$, $\dot Y_1(0)=8$.}.
The solutions  provide qualitative information  about product dynamics for two economies influencing each other. The particular time behavior of products $Y_1,\,Y_2$  obviously depends on the particular form of the constraint. Note that, for realistic modeling, the particular coefficients $k,\,\alpha,\,\beta$ should be estimated from  observed data.
Since the qualitative results of the test case  are for guidance only, 
we can state that, although  being simple and based on an (old-fashioned) classical physically motivated model, the solution 
is in a  qualitative correspondence with the observed behavior of the true interacting economies. 
\begin{figure}[th]
\begin{center}
\scalebox{0.25}{\includegraphics{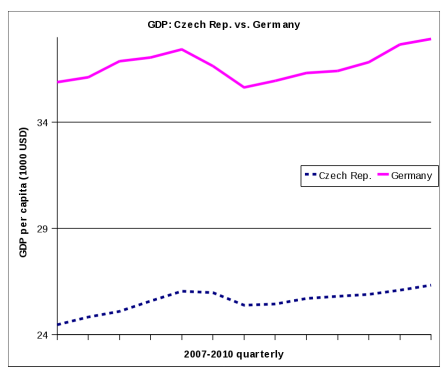}}
\caption{\label{GDP} GDP per capita in Germany and Czech Republic: 2007 to 2010, quarterly (source: http://stats.oecd.org, expenditure approach, current prices, seasonally adjusted)}
\end{center}
\end{figure} 
At Fig. \ref{GDP} we can see the GDP time series of Czech Republic and Germany in the period 2007-2010. As we have already mentioned there is a considerable trading between these two economies which possibly make a connection between the time development of their products  as qualitatively visible on  Fig. \ref{GDP}.  This connection could be qualitatively expressed by the constraint (\ref{vazba1}) and the model of two intercating economies  (\ref{redrov1}-\ref{redrov2})  could enable to describe the business cycles of the two economies.

Thus,  we have presented an example of how an economic model can be designed using the physical motivation. First the  classical model of product dynamics based on the mechanical model of an oscillator was employed to describe an isolated economy. 
Then this  model was extended to a model of two interacting economies using an oscillator-based approach, i.e. we treat each of the economies as a single  component of the system with two degrees of freedom. The existing  observed relationship between the economies was modeled by a constraint and the qualitative results of the model when compared to the  real data (see Fig. \ref{GDP}) are promising. In  economic science, such a model design could be successful after careful validation based on a concrete parameter estimation. 
As in other models, probably only some aspects of the observed behavior could be captured by the newly developed model and,  therefore, a debate about the relevance of the assumptions made and the methodology used might be  sparked. Nevertheless, such debates take place for any economic model no matter whether with a  physical or other background. 
An important fact is that, so far,  hardly any model has been   validated in economics in the sense of physics, since there always can be found significant discrepancies between the data predicted by the model and  those observed in the real economy world.  Indeed,  economic science is still trying to find the appropriate theories, methods and even the calculus to achieve the status of an unchallenged source of scholar knowledge.

\section{The economic growth model using calculus of variations}
While the short run variation in product (measured by gross domestic product -GDP- per capita ) is usually termed {\it business cycle}, the long run increase of per capita GDP is called {\it economic growth}. Economic growth is primarily driven by improvements in productivity, which involves producing more goods and services with the same inputs of labor, capital, energy and materials (therefore there is a lasting  demand  for  innovations and technological improvements in developed economies).  The long-run path of economic growth is one of the central questions of economics. An increase in GDP of a country greater than population growth is generally taken as an increase in the standard of living of its inhabitants, hence the aim of the policy makers is to achieve steady economic growth. To understand better the phenomena of economic growth and  to obtain a quantitative support for making  appropriate decisions a number of mathematical models of economic growth  was  developed  during past fifty years. Again- as we have already mentioned when generally speaking about economic models- all  models are representing the economic reality to individual extent and they differ in the assumptions made about how the economic characteristics are mutually interacting  and influencing the final growth. 

In this section, the  Cass-Koopmans-Ramsey model of economic growth   based on variational approach will be presented and  the methodology and motivations  behind an economic variational approach in comparison  to  traditional  usage of variational calculus in mechanics will be investigated. Since the aim of this section is not to provide the reader with a thorough economical background  of the model, we will describe  the philosophy of the model intuitively (for more detailed notes about particular assumptions and economical background see e.g. \cite{Romer}).

\subsection{The  Cass-Koopmans-Ramsey model}

This time, there are two "players"  in the  "economy game": firms and households (so the flows are similar to the previous model illustrated in Fig. \ref{picture-economy} except the missing government purchases). The aim of the model is to answer the question of how much the  product of economy at any point of time should be spent for immediate consumption to yield current utility, and how much of it should be  saved (and invested)\footnote[2]{In economics the  product (which is considered to be equal to the  income)  can either be consumed or saved, but  what is saved results in investment. For example we can imagine, that the household saves some amount of money in a bank. But the bank uses the money to provide the firms with loans and the firms use the extra money for investment purchases.} 
so as to enhance future production and consumption and, hence, yield future utility. The criterion of optimality is the social welfare which is given by social utility $U$. There are two   time-dependent  variables through which the desired optimal development of the economy can be attained:  consumption  $c(t)$ and capital  $k(t)$ (measured in  units advantageous for the model purposes, for details see \cite{Romer}). Consumption represents the purchases of goods and services by households  while capital is a factor of production, used to produce goods or services, that is not itself significantly consumed (though it may depreciate) in the production process (e.g. machinery, buildings, vehicles).  Typically, in economics there is considered a relationship between the investment and the time derivative of the  capital. In Cass-Koopmans-Ramsey model, the particular relationship takes the form

\begin{equation}
\dot k=f(k)-c-(n+g)k, \label{vazba2}
\end{equation}
where $f(k)$ denotes the product (more precisely it is a known  production function prescribing how the total production in economy is dependent on the capital available). Expression $f(k)-c$ has the meaning of  investment\footnote[4]{This expression  is connected with relationship $Y=C+S$ used in the previous example (see item (i)). Remember, that it means that the total income (or product, which is equivalent as we have already discussed) in economy is distributed among consumption and savings. Since what is saved is supposed to be invested, the savings define the investments in monetary units. Thus 

investments$\equiv$ savings = product- consumption.}  and $(n+g)k$  is the amount of investment that must be done just to keep $k$ at its existing level ($n$ and $g$ being  constants). The equation (\ref{vazba2}) states that the change in capital is equal to the investments less the replacement purchases (i.e. if we  buy new machinery only to replace the broken ones then, even we made  investments, we have not increased   capital- we still have the same production factors for our production process as before the investment purchases.)

The task in Cass-Koopmans-Ramsey growth model is to maximize social utility $U(c)$ (which  depends only  on consumption)\footnote[3]{It is typically assumed in economics that the utility of the society is given by the consumption of the people only. This assumption can be put into question and actually, several authors did so. It is  obvious that if the criterion of optimality in economy is given simply by consumption, then many  essential factors considering e.g. natural conditions can be omitted.} via time paths of variables $c$ and $k$ under the constraint (\ref{vazba2}). For the particular form of the utility function defined in Ramsey model (see \cite{Chiang} or \cite{Romer}),
we can write the overall optimization problem as:
\begin{eqnarray}
{\rm max}&&\hskip 1cm U=\int_0^\infty Be^{-\beta t}\frac{c^{1-\theta}}{1-\theta}{\rm d}t\label{integr}\\
&& s.t. \hskip 0.6cm f(k)-c-(n+g)k-\dot k=0,\label{vazba3}
\end{eqnarray}
where $B,\,\beta,\,\theta,\,n,\,g$ are positive constants with $0< \beta < 1$. The form of the consumption function is the matter of intensive investigation and the particular form used in the Cass-Koopmans-Ramsey model represents just one approach. The term $e^{-\beta t}$  ensures {\it discounting} of economical quantities. It means that the flows of particular quantity (social utility in our problem) to come have smaller weight that its current flow. Note that this approach was doubted by many authors because by this model the priority  is given to (our) current consumption at the expense of the consumption of future generations. 

From the mathematical point of view this is a constrained variational problem with a  velocity-dependent constraint (\ref{vazba3}). 
Solving (\ref{integr}-\ref {vazba3}), we obtain the optimal growth path, in other words,  we realize how $c$ and $k$ must behave over time to achieve the maximal lifetime utility from the consumption flow given the relation (\ref{vazba3}) between the product, consumption and investment. Note that $c$ in (\ref{integr}) could be substituted from (\ref{vazba3}) to obtain an unconstrained variational problem with the Lagrangian dependent on both  variable $k$ and its first derivative $\dot k$.
 Because the form (\ref{integr}-\ref{vazba3}) of the growth model is typically used  in economics (thanks  to its better economic interpretation  and reasoning), we will not "simplify" the initial problem (\ref{integr}-\ref{vazba3}) and it will be treated as a constrained variational problem for the solution of which  the vakonomic approach will be employed.    The Lagrangian for the constrained problem (\ref{integr}-\ref{vazba3})  takes the vakonomic  form


\begin{equation} L=Be^{-\beta t}\frac{c^{1-\theta}}{1-\theta}+
\lambda(f(k)-c-(n+g)k-\dot k),\end{equation}
were $\lambda=\lambda(t)$ is the Lagrange multiplier. Considering three variables $(c,\,k,\,\lambda)$  instead of the initial pair $(c,\,k)$, we obtain the variational dynamical equations:
\begin{eqnarray}
e^{-\beta  t}c^{-\theta}-\lambda&=&0,\label{EL1}\\
\lambda \frac{{\rm d}f}{{\rm d}k}-(n+g)\lambda+\dot\lambda&=&0,\label{EL2}\\
f(k)-c-(n+g)k-\dot k&=&0.\label{EL3}
\end{eqnarray}
Note that the vakonomic approach accounts for the  main technique used to solve variational constraint dynamical problems in economics (see \cite{Chiang},\cite{Romer},\cite{Aghion} for typical examples). For the description and some notes on the comparison of vakonomic and nonholonomic approach see Appendix 1.

\begin{figure}[th]
\begin{center}
\scalebox{0.6}{\includegraphics{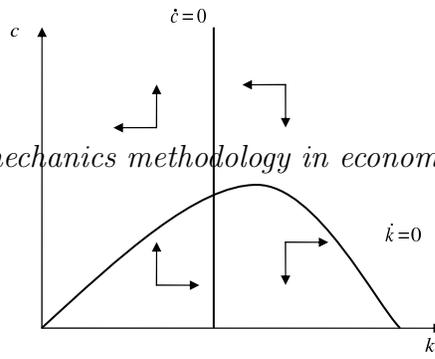}}
\caption{\label{Ramsey} The dynamics of $c$ and $k$ in Ramsey model}
\end{center}
\end{figure} 

\subsection{Results and Discussion}

  Expressing $\lambda$ from (\ref{EL2}) and substituting it into equation (\ref{EL1}), we arrive after some calculation  to a system  of two  equations  for two unknown variables $c(t),\,k(t)$: 
 \begin{eqnarray}
\frac{\dot c}{c}&=&\frac{\frac{{\rm d}f}{{\rm d}k}-\beta -n-g}{\theta}\label{Ra1}\\
\dot k&=&f(k)-c-(n+g)k.\label{Ra2}
\end{eqnarray}
Then, it is not difficult  to find a solution to our problem, i.e. to find an optimal consumption  and capital paths. 
Nevertheless, in economics  the functions are not normally given explicitly, but only assumed to have characteristic qualitative properties (such as  that the utility is an increasing non convex function of consumption). Thus the problems have an air of "theory" rather than "computation" and reaching a particular solution is neither possible nor necessary.  Yet, the content of the problems is meaningful and analyzing the qualitative  characteristics of the solutions often generates important insights into economic behavior. In models of economic dynamic optimization, two-variable  diagrams are prevalently employed to obtain these qualitative analytical results.

The behavior of an economy in the Cass-Koopmans-Ramsey problem is typically described in terms of the  evolution of $c$ and $k$ using a diagram  at Fig. \ref{Ramsey}. 
This diagram reflects the dynamical equations  (\ref{Ra1}-\ref{Ra2}) where the arrows show the directions of change of both $c$ and $k$.  
  In mechanics, we can plot similar diagrams but this is mostly done for a particular solution. 	In such a case the trajectory of the system in the  configuration space is obtained. The diagram in Fig.  \ref{Ramsey} is used to identify the qualitative optimal behavior of an economy given the initial conditions and additional requirements (such as  $k\geq 0$). Then, the result consists  in a  verbal description of the optimal behavior rather then in an analytical solution. As an example, let us discuss the top left "quadrant" of the graph. The consumption in economy is high and rising and $k$ eventually reaches zero. When $c$ continues to rise  $k$ must become negative. But this cannot occur. Since the product is zero when $k$ is zero, the  consumption must drop to zero. 
Therefore, such paths can be ruled out from consideration about realistic behavior of an economy. In a similar way,  other "quadrants" in Fig. \ref{Ramsey} can be analyzed and we would arrive to a conclusion that the possible time path of the system is the one driving into the point where $\dot k=0,\, \dot c =0$. Then the economy is said to "move along the saddle path" to the equilibrium  in $\dot k=0,\, \dot c =0$.  The equilibrium is Pareto-efficient which means  that it is impossible to make anyone better off without making someone else worse off. The explanation lies in the fact  that all the households in the model  have the same utility. The equilibrium produces  the highest possible utility among allocations of capital and consumption  that treat all households in the same way. Once the economy arrives to this equilibrium it can not do better  and therefore we do not change the level of capital and consumption anymore, i.e. $\dot k=0,\, \dot c =0$  (for details  see  \cite{Romer}).

We can see clearly the difference between the  application of variational calculus in mechanics  where we are searching for a particular solution which  prescribes the time evolution of the system precisely, and the economic application, where  the behavior of the system is  only qualitatively analyzed.

\subsection{ Lagrange multiplier interpretation}
A first look at a nontrivial constrained optimization economic problem may be surprising for the student of physics. Leaving aside the  unusual definition of  feasible region in economic optimization problem (mostly  restricted to non-negative values of variables), the economic interpretation of the Lagrange multiplier is worth a discussion. Mathematical texts provide no interpretation of the Lagrange multiplier $\lambda$, leaving the student with the impression that $\lambda$ has no significance beyond providing an extra variable which  transforms a constrained problem into an unconstrained, higher dimensional one. But in economic problems,  the Lagrange multiplier can usually be interpreted as the rate of change of optimal value of the criterion function relative to some parameter.

The constraint  (\ref{vazba3}) describes the  capital accumulation and thus $\lambda$ should correspond  to  the value of having a tiny bit more capital (see Appendix 2 for  simple explanatory example). Hence, we can find the meaning of the Lagrange multiplier by answering the question of how the unit change in capital available in economics will affect the  utility  in optimum. Let us derive the influence of change in capital onto a "present value" instantaneous utility $\bar u= u(c) e^{-\rho t}$ (see (\ref{integr})):
\begin{equation}
\frac{{\rm d} \bar u}{{\rm d} k}=\frac{\partial \bar u}{\partial k}+\frac{\partial \bar u}{\partial c}\frac{\partial c}{\partial k}.\label{LM} 
\end{equation}
From the first variational dynamical  equation (\ref{EL1}) we obtain
\begin{equation}
\lambda=\frac{\partial \bar u}{\partial c},\label{LM1}
\end{equation}
and differentiating the constraint (\ref{EL3}) we get
\begin{equation}
\frac{\partial c}{\partial k}=\frac{\partial f}{\partial k}-(n+g).\label{LM2}
\end{equation}
Substituing into relation (\ref{LM}) from (\ref{LM1}-\ref{LM2}) and having on mind that $\frac{\partial\bar u}{\partial k}=0$, we obtain
\begin{equation}
\frac{{\rm d} \bar u}{{\rm d} k}=\lambda\left(\frac{\partial f}{\partial k}-(n+g)\right). 
\end{equation}
That is, the extra unit of capital will raise the flow of output by an amount $\partial f / \partial k$ each unit of which (without the unit replacement production $n-g$ needed for keeping the capital at existing level and thus not intended for consumption) has a utility value  of $\lambda$. Multiplier $\lambda$ is referred to as {\it a shadow value of capital} (evaluated in present value).
New economical interpretation arises then from  variational equation (\ref{EL1}) and (\ref{LM1}), respectively: 
It states that the marginal  utility of consumption\footnote{utility gained (or lost) from an increase (or decrease) in the consumption }   is equal
to the shadow value of capital, $\lambda$. Thus, at the optimum the consumer is indifferent between
consuming an additional unit and investing it. Remember, that the consumer  must decide whether  to consume  or to save (which directly generates the investments) additional unit  (of income).  If it is consumed it brings direct utility, if it is invested (into capital) then it increases the product and future consumption. 
 If the marginal utility of
consumption was larger than the shadow value of capital,
\begin{equation}
\frac{\partial \bar u}{\partial c}> \lambda,
\end{equation} then capital would
be too high and consuming more and saving less would increase the utility.
Similarly, if the marginal utility of consumption was lower that the shadow value of capital
\begin{equation}
\frac{\partial \bar u}{\partial c}<\lambda,
\end{equation}  then capital is too low and
 the households should save more to increase utility through higher future
consumption.
This example of Lagrange multiplier  interpretation together with the subsequent discussion of  dynamic equations represent the typical approach of optimization problem analysis in economics.

In mechanics, no  interpretation is used for multipliers in optimization problems, but remember that the multipliers  arise in the constrained forces, which are important from the physical point of view. The concept of constrained forces, on the other hand, has no significance in economic problems although the multipliers do. Hence, although the computational routine remains similar, there are distinct interpretations of the tools used.

\section{Conclusion}

Classical mechanics has played a significant role in the development of economic thinking  influencing it  in both principles and the calculus. Currently, different attitudes to the adoption of physical methodology for economic purposes can be distinguished. A strong one supports the philosophy that applying the well developed physical methods could quickly provide the economists  with working and applicable models. This strong motion in economic science is reflected by a newly arisen term "econophysics" which currently designates one of the possible fields where the undergraduates and graduates of physics could be involved. 

The examples presented in the paper enable the undergraduates to meet for the first time this highly modern and progressively expanding field, and to get in touch with different philosophy of building scholar knowledge in economics. These examples  may serve as a motivation  for  students to further study the economics - mechanics (or more generally physics) intersections. In addition, studying these problems makes the undergraduate students of physics face the  different usage of a known methodology: the constraint variational calculus appears to be a flexible approach which in each discipline provides  specific information and the nonholonomic dynamics of systems, based on constraint forces, could be applied even outside mechanics.  

In physics education, the economic examples presented  can serve not only as a demonstration of interdisciplinary applications of methods typically used in mechanics,  but  can also  provide the teachers with an aid for demonstrating what are general mathematical and what are specifically physical features of  the mechanics methodology. In this way, the students can achieve a deeper insight into the physical background of what they have learned in classical mechanics courses.






\begin{ack}
\noindent The research is supported by the grant MSM6215648904 of the Ministry of Education, Youth and Sports of the Czech Republic. 
\end{ack}

\newpage

{\bf Appendix 1}

Consider a variational  system  with Lagrangian $L=L(t,q,\dot q)$, which is  subject to constraint $f(t,q,\dot q)=0$. 
The { \it nonholonomic approach } consists in incorporating the constraint forces into the Lagrange equations:
\begin{eqnarray} 
\frac{\partial L}{\partial q}  -  \frac{{\rm d}}{{\rm d }t}\frac{\partial L}{\partial \dot q}= \left(\frac{\partial f}{\partial \dot q}\right)^T\cdot \mu, \label{nonh1}\\
f(t,q,\dot q)=0, \label{nonh2}
\end{eqnarray}
where $\mu$ are the Lagrange multipliers.  The nonholonomic approach is typically used for solving the mechanical systems with nonholonomic constraints although in mechanics the non-holonomic constraints are frequently affine in velocities.
The form of constraint forces for classical ideal constraints is a consequence of d'Alemberts principle.
Note, that the nonholonomic approach has been extended also to more general constraints by the work of Chetaev \cite{Chetaev} and others (see e.g. \cite{Leon}; for  instructive  nonholonomic mechanical problems see e.g. \cite{valce}, \cite{valce2}, \cite{snakeboard}).

The mathematical concept of nonholonomic dynamics had been  kept  essentially unchanged until  30 years ago when a new dynamics of velocity constrained mechanics system was introduced by Kozlov \cite{Kozlov}. This new mechanics was called {\it vakonomic} being "variational axiomatic kind". If we adopt a variational approach by requiring the motion to be a stationary curve of the action functional among all curves having the same end points and satisfying the nonholonomic constraints, then we get a vakonomic motion, i.e. we search for the solution of (unconstrained) variational problem associated to the Lagrangian function
$$\bar L(t,q,\dot q,\lambda, \dot \lambda)=L(t,q,\dot q)+ \lambda f(t,q, \dot q).$$
Namely, the vakonomic motions can be obtained by the Lagrange equations
\begin{eqnarray} 
\frac{\partial \bar L}{\partial q}  -  \frac{{\rm d}}{{\rm d }t}\frac{\partial \bar L}{\partial \dot q}= 0,\\
\frac{\partial \bar L}{\partial\lambda}  -  \frac{{\rm d}}{{\rm d }t}\frac{\partial \bar L}{\partial \dot \lambda}= 0,
\end{eqnarray}
which give
\begin{eqnarray} 
\frac{\partial  L}{\partial q}  -  \frac{{\rm d}}{{\rm d }t}\frac{\partial  L}{\partial \dot q} -\left(\frac{\partial f}{\partial \dot q}\right)^T\cdot\dot \lambda +\left(\frac{\partial f}{\partial q}  -  \frac{{\rm d}}{{\rm d }t}\frac{\partial f}{\partial \dot q}\right)^T\cdot \lambda= 0,\label{vak1}\\
f(t,q,\dot q)=0. 
\end{eqnarray}
Comparing the nonholonomic (\ref{nonh1}) and vakonomic equations (\ref{vak1}) we arrive to  a  conclusion that the arising systems of differential equations are not equivalent unless 
$$\frac{\partial f}{\partial q}  -\frac{{\rm d}}{{\rm d }t}\frac{\partial f}{\partial \dot q}=0.$$
The author himself  in \cite{Kozlov} says: "...vakonomic dynamics, which is an internally consistent model that can be applied to description of the motion of any mechanical systems, is as "true" as traditional nonholonomic mechanics. The issue of the choice of model for each particular case is ultimately resolved by experiment."  While there are doubts that vakonomic dynamics is a satisfactory model for nonholonomic systems in mechanics (see e.g. \cite{Zampieri}), in economics vakonomic approach is the one typically used for the  solution of dynamical optimization problems with velocity dependent constraints. Intuitively, the  nonholonomic concept based on the  definition of constrained forces  developed for mechanics could not be valid for the economic problems unless the background of the model is physically motivated.  The vakonomic approach, if chosen for solving the constrained systems in economics, can be justified simply by the results it provides (for more examples on vakonomic approach in economics see e.g. \cite{Grubbstrom}, \cite{Molinder}).

\vskip\baselineskip
\vskip\baselineskip

{\bf Appendix 2}

In  constrained optimization in economics, the value of the Lagrange multiplier at the optimal solution is referred to as a  {\it shadow price}.  Shadow price is the change in the objective value of the optimal solution  obtained by relaxing the constraint by one unit.

 Consider a simple (static) business optimization problem of profit maximization in a firm. Assume that there are only two products and the firm is deciding about the amount of these two products $x_1,\,x_2$ to be produced in next period. The   profit and  number of working hours needed   per unit of each product are known and denoted by $a_1,\,a_2$ and  $l_1,\,l_2$, respectively. The operating time limit for next period is $B$ (for example, we have B working hours available on certain machine for next month). We get the linear programming problem:
\begin{eqnarray}
{\rm max}&&\hskip 1cm  \pi =a_1x_1+a_2x_2\label{UF}\\
&& {\rm s. t.} \hskip 0.6cm l_1x_1+l_2x_2=B.\label{B}
\end{eqnarray}
We introduce a  multiplier $\lambda $ and form the  Lagrangian
$$L=\pi+\lambda(B-l_1x_1-l_2x_2).$$
Assuming that the firm maximizes profit (given the constraint (\ref{B})), the optimal quantities $x_1^\star,\,x_2^\star$ and the multiplier $\lambda^\star$ necessarily satisfy the first-order conditions:
\begin{eqnarray}
\frac{\partial L}{\partial x_1}&=&\frac{\partial \pi}{\partial x_1}-\lambda l_1=0, \label{1rada} \\
\frac{\partial L}{\partial x_2}&=&\frac{\partial \pi}{\partial x_2}-\lambda l_2=0, \label{1radb}\\
\frac{\partial L}{\partial\lambda}&=&B-l_1 x_1-l_2 x_2=0, \label{1radc}
\end{eqnarray}
and differentiation of constraint (\ref{B}) yields
\begin{equation}
l_1\frac{\partial x_1}{\partial B}+l_2\frac{\partial x_2}{\partial B}=1.\label{pomvazB}
\end{equation}
Now, using the chain rule  and equations (\ref{1rada}), (\ref{1radb}), (\ref{pomvazB}) we obtain
\begin{equation}
\frac{\partial\pi}{\partial B} = \frac{\partial \pi}{\partial x_1} \frac{\partial x_1}{\partial B}+\frac{\partial \pi}{\partial x_2} \frac{\partial x_2}{\partial B}=\lambda l_1\frac{\partial x_1}{\partial B}+\lambda l_2\frac{\partial x_2}{\partial B}=\lambda. 
\end{equation}
Hence, the Lagrange multiplier $\lambda$  measures how the total profit responds to the unit change in total machine operating time available. The shadow price is here the maximum price the manager would be willing to pay for operating the production line for an additional unit of time, based on the benefits he would get from this change. For example, if $\lambda=1 $ Euro, then running the machine and producing for an additional hour will gain the profit of $1 $ Euro and the manager should not pay more for this additional working hour than is the value it produces (for more see e.g. \cite{Baxley}). 

\vskip\baselineskip

\end{document}